\documentclass[useAMS,usenatbib,aas_macros]{mnras}
\usepackage{graphicx}
\usepackage{epstopdf}
\usepackage{bm}
\usepackage{times}
\usepackage{amsmath,amssymb}
\usepackage{slashed}
\usepackage{color}
\usepackage{natbib}
\usepackage{gensymb}
\usepackage{hyperref}
\usepackage[]{appendix}
\usepackage{multirow}
\usepackage{tabularx}

\usepackage{eso-pic}% http://ctan.org/pkg/eso-pic

\AddToShipoutPictureBG*{%
  \AtPageUpperLeft{%
    \hspace{17cm}%
    \raisebox{-3.5\baselineskip}{%
      \makebox[0pt][l]{\textnormal{DES 2016-0158}}
}}}%

\AddToShipoutPictureBG*{%
  \AtPageUpperLeft{%
    \hspace{0.75\paperwidth}%
    \raisebox{-4.5\baselineskip}{%
      \makebox[0pt][l]{\textnormal{FERMILAB PUB-16-231-AE}}
}}}%

\begin{document}

\title[AGN in DES Clusters]{The Evolution of Active Galactic Nuclei in Clusters of Galaxies from the Dark Energy Survey}

\author[Bufanda et al.]{
\parbox{\textwidth}{
\Large
E.~Bufanda$^{1}$,
D.~Hollowood$^{1}$,
T.E.~Jeltema$^{1}$,
E.~S.~Rykoff$^{2,3}$,
E.~Rozo$^{4}$,
P.~Martini$^{5,6}$,
T. M. C.~Abbott$^{7}$,
F.~B.~Abdalla$^{8,9}$,
S.~Allam$^{10}$,
M.~Banerji$^{11,12}$,
A.~Benoit-L{\'e}vy$^{13,8,14}$,
E.~Bertin$^{13,14}$,
D.~Brooks$^{8}$,
A. Carnero Rosell$^{15,16}$,
M.~Carrasco~Kind$^{17,18}$,
J.~Carretero$^{19,20}$,
C.~E.~Cunha$^{2}$,
L.~N.~da Costa$^{15,16}$,
S.~Desai$^{21,22}$,
H.~T.~Diehl$^{10}$,
J.~P.~Dietrich$^{21,22}$,
A.~E.~Evrard$^{23,24}$,
A.~Fausti Neto$^{15}$,
B.~Flaugher$^{10}$,
J.~Frieman$^{10,25}$,
D.~W.~Gerdes$^{24}$,
D.~A.~Goldstein$^{26,27}$,
D.~Gruen$^{2,3}$,
R.~A.~Gruendl$^{17,18}$,
G.~Gutierrez$^{10}$,
K.~Honscheid$^{5,28}$,
D.~J.~James$^{7}$,
K.~Kuehn$^{29}$,
N.~Kuropatkin$^{10}$,
M.~Lima$^{30,15}$,
M.~A.~G.~Maia$^{15,16}$,
J.~L.~Marshall$^{31}$,
P.~Melchior$^{32}$,
R.~Miquel$^{33,20}$,
J.~J.~Mohr$^{21,22,34}$,
R.~Ogando$^{15,16}$,
A.~A.~Plazas$^{35}$,
A.~K.~Romer$^{36}$,
P.~Rooney$^{36}$,
E.~Sanchez$^{37}$,
B.~Santiago$^{38,15}$,
V.~Scarpine$^{10}$,
I.~Sevilla-Noarbe$^{37}$,
R.~C.~Smith$^{7}$,
M.~Soares-Santos$^{10}$,
F.~Sobreira$^{39,15}$,
E.~Suchyta$^{40}$,
G.~Tarle$^{24}$,
D.~Thomas$^{41}$,
D.~L.~Tucker$^{10}$,
A.~R.~Walker$^{7}$
\begin{center} (The DES Collaboration) \end{center}
}
\vspace{0.4cm}
\\
\parbox{\textwidth}{
%\scriptsize
$^{1}$ Department of Physics and Santa Cruz Institute for Particle Physics, University of California, Santa Cruz, CA 95064, USA\\
$^{2}$ Kavli Institute for Particle Astrophysics \& Cosmology, P. O. Box 2450, Stanford University, Stanford, CA 94305, USA\\
$^{3}$ SLAC National Accelerator Laboratory, Menlo Park, CA 94025, USA\\
$^{4}$ Department of Physics, University of Arizona, Tucson, AZ 85721, USA\\
$^{5}$ Center for Cosmology and Astro-Particle Physics, The Ohio State University, Columbus, OH 43210, USA\\
$^{6}$ Department of Astronomy, The Ohio State University, Columbus, OH 43210, USA\\
$^{7}$ Cerro Tololo Inter-American Observatory, National Optical Astronomy Observatory, Casilla 603, La Serena, Chile\\
$^{8}$ Department of Physics \& Astronomy, University College London, Gower Street, London, WC1E 6BT, UK\\
$^{9}$ Department of Physics and Electronics, Rhodes University, PO Box 94, Grahamstown, 6140, South Africa\\
$^{10}$ Fermi National Accelerator Laboratory, P. O. Box 500, Batavia, IL 60510, USA\\
$^{11}$ Institute of Astronomy, University of Cambridge, Madingley Road, Cambridge CB3 0HA, UK\\
$^{12}$ Kavli Institute for Cosmology, University of Cambridge, Madingley Road, Cambridge CB3 0HA, UK\\
$^{13}$ CNRS, UMR 7095, Institut d'Astrophysique de Paris, F-75014, Paris, France\\
$^{14}$ Sorbonne Universit\'es, UPMC Univ Paris 06, UMR 7095, Institut d'Astrophysique de Paris, F-75014, Paris, France\\
$^{15}$ Laborat\'orio Interinstitucional de e-Astronomia - LIneA, Rua Gal. Jos\'e Cristino 77, Rio de Janeiro, RJ - 20921-400, Brazil\\
$^{16}$ Observat\'orio Nacional, Rua Gal. Jos\'e Cristino 77, Rio de Janeiro, RJ - 20921-400, Brazil\\
$^{17}$ Department of Astronomy, University of Illinois, 1002 W. Green Street, Urbana, IL 61801, USA\\
$^{18}$ National Center for Supercomputing Applications, 1205 West Clark St., Urbana, IL 61801, USA\\
$^{19}$ Institut de Ci\`encies de l'Espai, IEEC-CSIC, Campus UAB, Carrer de Can Magrans, s/n,  08193 Bellaterra, Barcelona, Spain\\
$^{20}$ Institut de F\'{\i}sica d'Altes Energies (IFAE), The Barcelona Institute of Science and Technology, Campus UAB, 08193 Bellaterra (Barcelona) Spain\\
$^{21}$ Faculty of Physics, Ludwig-Maximilians-Universit\"at, Scheinerstr. 1, 81679 Munich, Germany\\
$^{22}$ Excellence Cluster Universe, Boltzmannstr.\ 2, 85748 Garching, Germany\\
$^{23}$ Department of Astronomy, University of Michigan, Ann Arbor, MI 48109, USA\\
$^{24}$ Department of Physics, University of Michigan, Ann Arbor, MI 48109, USA\\
$^{25}$ Kavli Institute for Cosmological Physics, University of Chicago, Chicago, IL 60637, USA\\
$^{26}$ Department of Astronomy, University of California, Berkeley,  501 Campbell Hall, Berkeley, CA 94720, USA\\
$^{27}$ Lawrence Berkeley National Laboratory, 1 Cyclotron Road, Berkeley, CA 94720, USA\\
$^{28}$ Department of Physics, The Ohio State University, Columbus, OH 43210, USA\\
$^{29}$ Australian Astronomical Observatory, North Ryde, NSW 2113, Australia\\
$^{30}$ Departamento de F\'{\i}sica Matem\'atica,  Instituto de F\'{\i}sica, Universidade de S\~ao Paulo,  CP 66318, CEP 05314-970, S\~ao Paulo, SP,  Brazil\\
$^{31}$ George P. and Cynthia Woods Mitchell Institute for Fundamental Physics and Astronomy, and Department of Physics and Astronomy, Texas A\&M University, College Station, TX 77843,  USA\\
$^{32}$ Department of Astrophysical Sciences, Princeton University, Peyton Hall, Princeton, NJ 08544, USA\\
$^{33}$ Instituci\'o Catalana de Recerca i Estudis Avan\c{c}ats, E-08010 Barcelona, Spain\\
$^{34}$ Max Planck Institute for Extraterrestrial Physics, Giessenbachstrasse, 85748 Garching, Germany\\
$^{35}$ Jet Propulsion Laboratory, California Institute of Technology, 4800 Oak Grove Dr., Pasadena, CA 91109, USA\\
$^{36}$ Department of Physics and Astronomy, Pevensey Building, University of Sussex, Brighton, BN1 9QH, UK\\
$^{37}$ Centro de Investigaciones Energ\'eticas, Medioambientales y Tecnol\'ogicas (CIEMAT), Madrid, Spain\\
$^{38}$ Instituto de F\'\i sica, UFRGS, Caixa Postal 15051, Porto Alegre, RS - 91501-970, Brazil\\
$^{39}$ ICTP South American Institute for Fundamental Research\\ Instituto de F\'{\i}sica Te\'orica, Universidade Estadual Paulista, S\~ao Paulo, Brazil\\
$^{40}$ Department of Physics and Astronomy, University of Pennsylvania, Philadelphia, PA 19104, USA\\
$^{41}$ Institute of Cosmology \& Gravitation, University of Portsmouth, Portsmouth, PO1 3FX, UK\\
}
}

\maketitle

\begin{abstract}

The correlation between active galactic nuclei (AGN) and environment provides important clues to AGN fueling and the relationship of black hole growth to galaxy evolution.  In this paper, we analyze the fraction of galaxies in clusters hosting AGN as a function of redshift and cluster richness for X-ray detected AGN associated with clusters of galaxies in Dark Energy Survey (DES) Science Verification data. The present sample includes 33 AGN with $L_X > 10^{43}$ ergs s$^{-1}$  in non-central, host galaxies with luminosity greater than $0.5 L_{\ast}$ from a total sample of 432 clusters in the redshift range of $0.1<z<0.95$.  Analysis of the present sample reveals that the AGN fraction in red-sequence cluster members has a strong positive correlation with redshift such that the AGN fraction increases by a factor of $\sim 8$ from low to high redshift, and the fraction of cluster galaxies hosting AGN at high redshifts is greater than the low-redshift fraction at $3.6\sigma$. In particular, the AGN fraction increases steeply at the highest redshifts in our sample at $z>0.7$. This result is in good agreement with previous work and parallels the increase in star formation in cluster galaxies over the same redshift range.  However, the AGN fraction in clusters is observed to have no significant correlation with cluster mass. Future analyses with DES Year 1 through Year 3 data will be able to clarify whether AGN activity is correlated to cluster mass and will tightly constrain the relationship between cluster AGN populations and redshift.

\end{abstract}


\section{Introduction}
Galaxies are a key component in the formation of large scale structure; however, their evolution and the mechanisms driving it on large time scales are not well-understood.  All massive galaxies are thought to host a supermassive black hole (SMBH) \citep[e.g.][]{1998IAUS..184..377F}, and observational studies have shown that the size of this SMBH is correlated with various properties of its host galaxy, such as the velocity dispersion of the central bulge \citep[e.g.][]{2000ApJ...539L...9F, 2000ApJ...539L..13G, 2000MNRAS.318L..35H, 2002ApJ...574..740T, 2016ASSL..418..263G} and central bulge formation \citep[e.g.][]{2007ApJ...664..198X, 2016ASSL..418..263G}. Hence, one can infer that the evolution of the central SMBH and its host galaxy are intimately connected. 

Black holes at the centers of galaxies become active when there is a large influx of gas onto the black hole that could arise from major mergers \citep[e.g.][]{1988ApJ...325...74S, 1991ApJ...370L..65B, 2006ApJS..163....1H}, particularly for the most luminous active galactic nuclei (AGN).  Lower luminosity AGN may instead be fed by minor mergers, recycled stellar material, or bars \citep[e.g.][]{1980ApJ...237..404S, 1988A&A...203..259N, 2008ApJ...687...59G, 2009MNRAS.397..623G, 2012ApJ...746..155R, 2014ApJ...783...40G}. When there is no more gas to accrete, the black hole returns to its dormant state. Thus, it is theorized that AGN are just a phase of evolution that every galaxy experiences.  Any mechanism that adds to the inflow of cold gas into a galaxy has the potential to ignite star formation as well as AGN activity. In fact, there is a correlation between AGN activity and star formation in galaxies and the two show similar cosmic evolution \citep[e.g.][]{1990MNRAS.242..271T, 1999MNRAS.310L...5F, 2003MNRAS.346.1055K, 2012ApJ...746..155R, 2014ApJ...783...40G}. AGN feedback is then thought to regulate galactic star formation, stripping the galaxy of the conditions necessary for stellar nurseries to form \citep[e.g.][]{1998A&A...331L...1S, 2005Natur.433..604D, 2005ApJ...630..705H, 2013ApJ...779L..13C, 2013ApJ...763...59R}. 

Galaxies in clusters are known to evolve at different rates than galaxies in the field. For example, in the local Universe the bulk of galaxies in clusters exhibit little or no star formation compared to galaxies in the field; however, the fraction of star-forming cluster galaxies increases strongly with redshift \citep[e.g.][]{1984ApJ...285..426B, 2005MNRAS.358...88W, 2006ApJ...642..188P, 2008ApJ...685L.113S, 2009ApJ...704..126H}.  In the dense environments of clusters and groups different physical processes may influence galaxy evolution including galaxy-galaxy mergers and harassment \citep[e.g.][]{1976ApJ...204..642R, 1980ApJ...236...43A, 1985MNRAS.215..517B, 1996Natur.379..613M, 2009ApJ...699.1595P}, strangulation \citep[e.g.][]{1980ApJ...237..692L, 2000ApJ...540..113B, 2008ApJ...672L.103K, 2008MNRAS.387...79V}, ram-pressure stripping \citep[e.g.][]{1972ApJ...176....1G, 1999MNRAS.308..947A, 2008MNRAS.383..593M} and evaporation by the hot intracluster medium \citep[][]{1977Natur.266..501C}.  These processes can hinder the availability and transport of cold gas to the galaxy. By extension, the lack of cold gas available to the galaxy affects AGN activity and evolution.  In addition, galaxies in clusters appear to have formed on average earlier than similar mass galaxies in the field \citep[e.g.][]{1996MNRAS.281..985V, 1997ApJ...478L..13K}.  Hence, the dependence of AGN activity on environment probes mechanisms driving AGN evolution as well as the correlation of AGN to star formation and galaxy growth which in turn depend on environment.

Studies have shown that at low redshifts, the fraction of galaxies in clusters hosting X-ray luminous AGN is lower than for field galaxies while the cluster AGN fraction is observed to be the same as the field AGN fraction at higher redshifts ($z\sim1$) \citep{2007ApJ...664L...9E, 2009ApJ...701...66M, 2010ApJ...714L.181K, 2013ApJ...768....1M}. Therefore, the cluster AGN fraction evolves more strongly leading to parity with field AGN at high redshift. It follows that a strong positive correlation between cluster AGN fraction and redshift is observed.  This increase in cluster AGN activity with redshift parallels the increase in the fraction of star forming galaxies in groups and clusters over a similar redshift range \citep[e.g.][]{1984ApJ...285..426B, 2005MNRAS.358...88W, 2006ApJ...642..188P, 2008ApJ...685L.113S, 2009ApJ...704..126H, 2014MNRAS.445.2725E}.  Similarly, \cite{2016ApJ...825...72A} find that the AGN fraction, with AGN selected based on their optical-IR spectral energy distributions (SEDs), increases with cluster redshift out to $z\sim2$ with the cluster AGN fraction exceeding the field fraction at $z>1$; the increase in AGN fraction mirrors an increase in star formation with cluster redshift also found by \cite{2016ApJ...825...72A}.  In addition, several studies have found that the AGN fraction in clusters increases with increasing distance from the cluster center with AGN fractions suppressed in cluster cores but reaching parity with the field at larger radii \citep{2009ApJ...701...66M, 2013MNRAS.429.1827P, 2014MNRAS.437.1942E, 2016MNRAS.461.2115D}.  A similar radial trend is also seen for star-forming galaxies in clusters; however, the detailed relationship between AGN and host galaxy properties appears to depend both on environment and on AGN selection method \citep[e.g.][]{2011ApJ...729...22A, 2012MNRAS.425.1215K, 2014MNRAS.442..314K}.  AGN host galaxies have also been found to have larger velocity dispersions than typical cluster members indicative of an infalling population \citep{2012ApJ...754...97H}, and \cite{2016A&A...592A..11K} find tentative evidence for an overdensity of X-ray AGN in superclusters, though their sample size is small.  

Galaxies in groups are more likely to interact with one another, which could ignite AGN through mergers and close interactions \citep{2007ApJ...654L.115S, 2008ApJ...682..803S}. In contrast, galaxies in dense cluster environments have velocity dispersions which are too high to allow bound pairs \citep{1998MNRAS.300..146G}. This, combined with the greater effects of cluster processes for higher mass clusters might imply that the AGN fraction should be lower in clusters compared to galaxy groups. In fact, several works find that the AGN fraction in galaxy groups is higher than the AGN fraction in clusters \citep{2008ApJ...682..803S, 2009ApJ...707.1691A, 2014ApJ...790...43O, 2014A&A...567A..83K, 2015MNRAS.446.2709E} though the trend is relatively mild.  In contrast, the fraction of star-forming galaxies in groups and clusters is comparable at low redshifts down to group masses of $\sim10^{13} M_{\odot}$ \citep[e.g.][]{2007ApJ...658..865J, 2010MNRAS.402L..59B, 2014MNRAS.445.2725E}, though this fraction appears to increase with redshift more strongly in lower mass systems \citep[e.g.][]{2006ApJ...642..188P, 2014MNRAS.445.2725E}.

In this paper we investigate the X-ray AGN fraction in red-sequence galaxy members of clusters of galaxies discovered in the Dark Energy Survey Science Verification data.  Clusters and their member galaxies are identified using the redMaPPer algorithm \citep{2014ApJ...785..104R, 2016arXiv160100621R} with redshifts ranging from 0.1 to 0.95. We cross-match AGN detected in {\em Chandra} observations overlapping the DES Science Verification fields to the cluster member galaxies identified by redMaPPer.  These data give us a large, uniformly selected sample of clusters with a wide range of optical richnesses and redshifts allowing us to probe AGN activity in clusters as a function of redshift and cluster mass. The outline of the paper is as follows: in \S2 and \S3 we present the selection of cluster galaxies, the X-ray analysis, and the AGN selection; \S4 describes the determination of the AGN fraction; \S5 the correlations of AGN fraction with cluster redshift and mass, and we summarize our results in \S6.

\section{Cluster Galaxy Sample and Selection}

The Dark Energy Survey (DES) is covering $\sim 5000$ deg$^2$ of the southern sky over a five year period in 5 bands ($grizY$) using DECam, a 570 megapixel imager \citep{1538-3881-150-5-150, 2016MNRAS.460.1270D}.  In this paper, we utilize clusters discovered in pre-survey Science Verification (SV)\footnote{https://des.ncsa.illinois.edu/releases/sva1} observations taken with DECam between November 2012 and February 2013 and processed by the DES data management system \citep{DESDM2012}.  The fields targeted in SV include a 139 deg$^2$ contiguous region in the South Pole Telescope East field \citep{2013ApJ...779...86S, 2015PhRvD..92b2006V, 2015PhRvL.115e1301C}, supernova fields, and three targeted massive clusters \citep{2015MNRAS.449.2219M}.  Clusters in this work were drawn from the $\sim$ 250 deg$^2$ SVA1 Gold data much of which is close to the expected full depth of the DES survey data.  These data are suitable for extragalactic science and exclude portions of the SV data south of $-61^{\circ}$ declination where proximity to the Large Magellanic Cloud leads to high stellar densities.

Clusters and their member galaxies were selected using the red-sequence Matched-Filter Probabilistic Percolation (redMaPPer) algorithm \citep{2014ApJ...785..104R}.  The DES SV redMaPPer cluster catalog is described in \cite{2016arXiv160100621R}.  As our study benefits from increased sample size and does not require the most conservative data quality cuts, we consider the expanded redMaPPer catalog as described in \cite{2016arXiv160100621R} which includes clusters with $0.1<z<0.95$ selected from a 208 deg$^2$ area after quality cuts.  The catalog includes more than 16,000 clusters with richnesses of $\lambda > 5$ and more than 1300 clusters with $\lambda > 20$.

redMaPPer assigns each galaxy in the vicinity of a cluster a probability of being a cluster member \citep{2014ApJ...785..104R}; the redMaPPer member catalog includes all galaxies with a probability of at least 1\% of being a cluster member with the corresponding cluster to which they likely belong \citep{2015MNRAS.453...38R}.  It should be noted that redMaPPer primarily selects red cluster galaxies as galaxies are required to lie on or near the cluster red sequence \citep{2015MNRAS.453...38R}; thus we are probing the fraction of red cluster galaxies hosting AGN.  Selection of blue cluster galaxies at the level of assigning individual galaxies cluster membership cannot be done robustly from photometry alone.  For high-luminosity AGN in particular ($L_X \gtrsim 10^{44}$ erg s$^{-1}$) the selection of red galaxies is expected to lead to some incompleteness in the AGN sample as their visible-wavelength counterparts are more likely to be blue \citep[e.g.][]{2013ApJ...768....1M}. 

In terms of aperture, redMaPPer determines a radial cut for detected clusters which scales with cluster richness, and the member catalog tabulates cluster member galaxies within this aperture.  In this way, richer, more massive clusters are properly assigned larger radii.  The radial cut and its scaling with richness are determined to minimize scatter in richness relative to other mass proxies \citep[see][]{2012ApJ...746..178R}; thus, this radius is not directly analogous to an overdensity radius such as $R_{200\textrm{m}}$\footnote{The radius at which the cluster density is 200 times the mean density of the universe at a given redshift}.  However, the redMaPPer radius, $R_{\lambda}$ does cover a significant fraction of $R_{200\textrm{m}}$ ranging from $R_{\lambda} \sim 0.5-2 R_{200\textrm{m}}$, depending on cluster richness and redshift (see the mass-richness relation of \citep{2016arXiv160306953S} and the definition of $R_{\lambda}$ in \citep{2012ApJ...746..178R}).

For our analysis, we consider all galaxies in the member catalog with luminosities greater than $0.5 L_{\ast}$ in z-band.  The member catalog extends to somewhat  lower luminosities, but this cut ensures that the galaxy catalog is complete out to the highest redshifts considered for our cluster sample.  Here $L_{\ast}(z)$ is calculated using a \cite{2003MNRAS.344.1000B} model to find the magnitude of a galaxy with a single burst of star formation at $z=3$ with solar metallicity and Salpeter IMF and normalized to match the $m_{\ast}(z)$ relation for the SDSS redMaPPer catalog at $z=0.2$ \citep{2016arXiv160100621R}.  As discussed in Section 4, the number of cluster AGN and the number of cluster galaxies sampled are properly modulated by their redMaPPer membership probabilities in determining the fraction of cluster galaxies hosting X-ray active AGN.  In total, more than 400 clusters and more than 6000 cluster galaxies brighter than $0.5 L_{\ast}$ fall within the Chandra observations used in our analysis.

\section{X-ray Data Reduction and Analysis}

We consider 103 archival {\em Chandra} ACIS-I or ACIS-S observations within the regions covered by DES SV\footnote{For a sense of the coverage and depth of archival Chandra data see \cite{2010ApJS..189...37E}.}.  The data were reduced using standard analysis procedures using CIAO version 4.7 \citep{2006SPIE.6270E..1VF}.  In brief, the data were reprocessed starting from the level 1 event file to apply the newest gain and charge transfer inefficiency (CTI) corrections and generate observation specific bad pixel files.  For observations taken in VFAINT mode, the background cleaning for very-faint mode data was applied.  Time periods contaminated by background flaring were detected and removed using the {\tt lc\_clean} routine.  We use {\tt wavdetect}, a wavelet source detection tool available in CIAO which accounts for the point spread function of off-axis sources to determine the number of point sources in the energy range 0.3--7.9 keV in each observation; these point source lists form the basis of our AGN candidates.

For each X-ray observation we cross matched detected X-ray point sources to DES redMaPPer cluster member galaxies using a 2" match radius\footnote{The astrometric accuracy of Chandra is typically better than 0.4 arcsec with nearly all sources localized to better than 1 arcsec (http://cxc.harvard.edu/proposer/POG/)}.  X-ray sources were checked visually to ensure that they were truly point-like rather than associated to the core of an X-ray bright cluster. This led to an initial sample of 160 X-ray point sources associated to cluster member galaxies.  We estimate the X-ray luminosity of each source in the rest-frame 0.3--7.9 keV band based on its observed count rate and assuming a power law spectrum with a photon index of $\Gamma=1.7$ and correcting for Galactic absorption; redshifts are taken to be the redMaPPer determined photometric redshift of the host cluster. Source count rates are determined in a circular region centered on the source with radius ranging from 4--25" depending on position on the detector to account for the larger Chandra PSF at larger off-axis angles. Background count rates are estimated using a local annular background region with inner radii matching the source outer radius and outer radii ranging from 8--40".  Sources with a signal-to-noise ratio below 3 were eliminated from our sample. To probe galaxies that undergo a similar evolution, we eliminated galaxies with and without AGN whose positions were located in the centers of their galaxy cluster as determined by redMaPPer.   In this way, we are probing AGN activity in satellite galaxies rather than central galaxies whose evolution is likely different.  Additionally, eliminating these sources concurrently reduces the chance that {\tt wavdetect} mistakenly detected the diffuse emission of the cluster core as a point source.  Outside of the central AGN, X-ray detected AGN in clusters have luminosities similar to local Seyfert and LINER galaxies.

To ensure uniform depth in the X-ray across the redshift range probed and given the varying depth of the archival X-ray observations, we made an additional cut on X-ray luminosity, limiting the sample to sources with $L_X > 10^{43}$ erg s$^{-1}$ (0.3--7.9 keV band, rest frame).  After making cuts on X-ray luminosity, host galaxy luminosity brighter than $0.5 L_{\ast}$ at the cluster redshift, eliminating central galaxies and those with low signal-to-noise, we obtained a final sample of 33 AGN detected in a sample of more than 5600 galaxies. The properties of redMaPPer cluster galaxies hosting the 33 AGN in our sample are listed in Table 1.  Of the original 160 X-ray point sources associated red-sequence cluster members, 45 are in galaxies which do not meet our z-band luminosity cut, 8 are central sources or associated to the redMaPPer central galaxy, 70 have $L_X < 10^{43}$ erg s$^{-1}$, and 4 have low signal-to-noise.  Below $z<0.4$, we are complete to X-ray AGN down to lower luminosities of $L_X > 10^{42}$ erg s$^{-1}$; for $0.1<z<0.4$ there are 15 additional AGN with $10^{42} < L_X <10^{43}$ erg s$^{-1}$ hosted by non-central galaxies of sufficient brightness to be included in our sample.  We will briefly discuss the AGN fraction including these sources when considering a possible trend in AGN fraction with richness, though we do not find that this changes our main conclusions.

\begin{table*}
\centering
    \begin{tabular}{| p{1.4cm} | p{1.4cm} | p{1.4cm} | p{1.4cm} | p{1.6cm} | p{1.4cm} | p{1.4cm} |p{1.4cm} |p{1.4cm} |} \hline
RA &	DEC&	Redshift ($z$)	&Cluster ID	&Richness ($\lambda$)	&P	&$P_{\textrm{free}}$	&Radius 	&$L_X$\\ 
(\degree) & (\degree) & & & & & &$(h^{-1}$ Mpc) &(erg s$^{-1}$) \\\hline
15.5915 &	$-$49.3596	&0.76	&16689	&6.2	&0.11	&1	&0.14	&2.62E+43\\
15.64846&	$-$49.4122	&0.77	&36215	&7.7	&0.02	&1	&0.58	&1.52E+43\\
35.99583&	$-$4.31834& 	0.92	&29626	&5.1&	0.07	&0.91	&0.50	&3.70E+43\\
36.33743&	$-$4.54087	&0.85	&12659	&10.5	&0.16	&1	&0.44	&4.40E+43\\
64.24188&	$-$47.8744& 	0.60&	313	&5.8	&0.13	&0.93	&0.12	&2.37E+43\\
66.4759&	$-$54.9158&	0.64&	20&	 90.1&	0.06&	1&	0.44&	3.34E+43\\
67.1305&	$-$53.8017&	0.28&	275&	 36.2&	0.23&	1&	0.60&	1.27E+43\\
70.4522&	$-$48.9130&	0.81&	45&	 86.8&	0.96&	1&	0.20&	2.02E+43\\
70.45632&	$-$48.8609&	0.79&	7417	& 7.8&	0.03&	0.98&	0.22&	2.26E+43\\
71.31361&	$-$58.818&	0.44&	24853&	5.4&	0.40&	1&	0.53&	1.07E+43\\
71.50422&	$-$58.78&	0.58&	4710	& 13.3&	0.58&	1&	0.26&	1.47E+43\\
71.6578&	$-$48.5762&	0.77&	353&	 47.2& 	0.48&	1&	0.55&	4.69E+43\\
71.6789&	$-$48.6007&	0.77&	353&	 47.2&	0.05&	1&	0.82&	1.60E+43\\
71.8516&	$-$58.9019&	0.68&	571&	 38.6&	0.05&	1&	0.75&	1.72E+43\\
77.3753&	$-$53.6786&	0.46&	269&	 54.6&	0.83	&1	&0.43	&1.29E+43\\
104.3816&	$-$55.9587&	0.63&	18380&	8.1&	0.26&	1&	0.68&	2.08E+43\\
104.515&	$-$56.0205&	0.30&	1&	 280.6&	0.70&	1&	1.14&	1.02E+43 \\
149.7007&	2.40267&	0.43&	10099&	12.2&	0.08&	1&	0.13&	2.73E+44\\
149.926&	2.52635&	0.71&	74&	 79.6&	0.97&	1&	0.05&	1.01E+43\\
149.9366&	2.67964&	0.91&	21429&	10.8&	0.01&	0.98&	0.50&	1.40E+44\\
149.9576&	2.39579&	0.940&	17879&	9.0&	0.17&	1&	0.36&	2.34E+43\\
150.0303&	2.35874&	0.78&	39122&	5.7&	0.73&	1&	0.11&	1.39E+43\\
150.0487&	2.32214&	0.77&	27256&	5.6&	0.06&	0.98&	0.36&	1.32E+43\\
150.1893&	2.60665&	0.91&	5322&	18.8&	0.02&	1&	0.55&	5.65E+43\\
150.1991&	2.59767&	0.91&	5322&	18.8&	0.57&	1&	0.73&	3.29E+43\\
150.4058&	2.51809&	0.89&	33128&	8.0&	0.01&	1&	0.15&	2.36E+43\\
150.4228&	2.12875&	0.91&	17188&	9.5&	0.06&	1&	0.61&	1.15E+43\\
150.4354&	2.1428&	0.91&	17188&	9.5&	0.03&	1&	0.29&	2.70E+43\\
150.504&	2.22447&	0.84&	9472&	15.1&	0.94&	1&	0.02&	5.85E+43\\
150.5376&	2.18815&	0.90&	16068&	5.9&	0.45&	0.99&	0.22&	3.72E+43\\
150.5404&	2.16791&	0.90&	16068&	5.9&	0.20&	0.63&	0.53&	2.76E+43\\
150.5731&	2.20353&	0.85&	12575&	11.4&	0.07&	1&	0.29&	1.78E+43\\
150.6348&	2.1657&	0.92&	40590&	7.2&	0.02&	1&	0.42&	5.58E+43\\
\hline
	\end{tabular}
	\caption{Properties of galaxies hosting X-ray detected AGN in the redMaPPer SVA1 expanded catalog which meet our X-ray and optical luminosity completeness cuts ($L_X > 10^{43}$ erg s$^{-1}$ and host galaxy luminosity brighter than $0.5 L_{\ast}$ in z-band).  Columns 3--5 give the redshift, ID, and richness of the host cluster from the redMaPPer catalog.  The photometric redshift uncertainty is $\sigma_z/(1+z) \sim 0.01$ over most of the redshift range probed rising to $\sim 0.02$ at the highest redshifts \citep{2016arXiv160100621R}. The product of columns 6 and 7 give the probability that the AGN host galaxy is a member of the cluster as discussed in \S 4.  Column 8 gives the projected radial distance of the host galaxy from the redMaPPer determined central cluster galaxy, and column 9 gives the X-ray luminosity assuming a power law spectrum with a slope $\Gamma=1.7$.}
	\label{agnlist}
\end{table*}

\section{AGN Fraction}

The AGN fraction is simply defined as the fraction of cluster galaxies which host X-ray active AGN.  In order to determine the AGN fraction, we first need to know how many cluster galaxies were actually sampled by the Chandra observations considered.  We also need to properly account for the probability that a given redMaPPer candidate cluster member galaxy is actually associated to a cluster.  Taking the redMaPPer catalog of cluster members within a radius $R_{\lambda}$ and brighter than $0.5 L_{\ast}$, we determined which galaxies fell within the field-of-view of an archival Chandra observation.  As the Chandra ACIS data is not of uniform depth across the field, we further examined the exposure maps for each observation to ensure that the depth at each galaxy position was sufficient to detect an AGN of $L_X = 10^{43}$ erg cm$^{-2}$ s$^{-1}$ at the redshift of the cluster.

Each galaxy is assigned a probability of being a cluster member by redMaPPer.  The total probability of a galaxy being a member of a given cluster is the product of the membership probability times the probability that the galaxy under consideration is not a member of a higher ranked cluster ($p_{\textrm{free}}$). We use this total probability, though in practice $>$75\% of our AGN host galaxies have $p_{\textrm{free}}=1$ and all but one have $p_{\textrm{free}}>0.9$.  The number of cluster AGN in a given redshift or richness bin is calculated as the sum of the membership probabilities of the AGN host galaxies; the fraction of cluster galaxies hosting AGN is then this sum divided by the sum of the membership probabilities of all galaxies within Chandra observations of sufficient depth in the same redshift or richness range.  The variances on the number of cluster AGN and number of total cluster galaxies are likewise calculated from the membership probabilities based on the variance of the Bernoulli distribution as
\[
\sigma^2 = \sum_{i=0}^{N} p_i(1-p_i)
\]
where the sum runs over all AGN host galaxies or all cluster galaxies in the sample under consideration, respectively, and these errors are then propagated into the error on the AGN fraction.  

\section{Results and Discussion}

\begin{figure}
\includegraphics[scale=0.5]{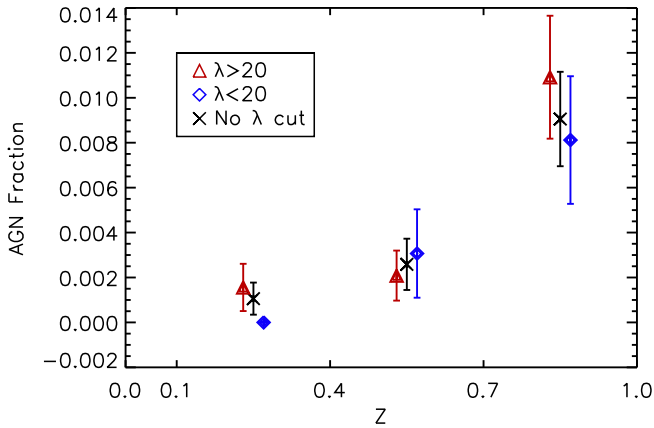}
\caption{Evolution of the AGN fraction in clusters from $z=0.1$ to $z=1.0$, all AGN with X-ray luminosity thresholds of $L_{X}>10^{43} erg s^{-1}$, in 3 redshift bins of $z<0.4, 0.4<z<0.7, 0.7<z<1.0$ , for AGN in high mass clusters ( $\lambda>20$, blue diamonds), AGN in low mass clusters ($\lambda<20$, red triangles), and AGN in clusters of all masses ( $\lambda > 5$, black crosses).  Points have been slightly offset in the x direction for visual clarity.}
\label{fig:withz}
\end{figure}

\begin{figure}
\includegraphics[scale=0.5]{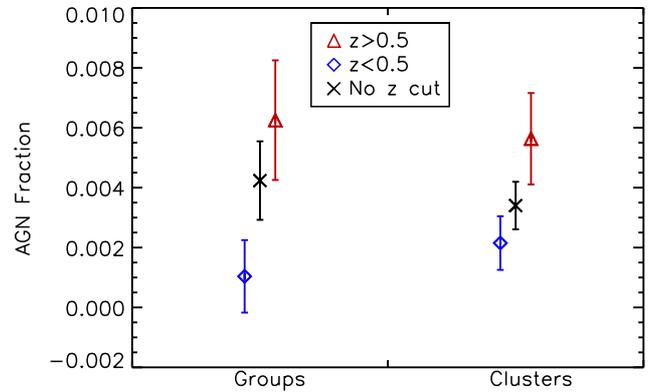}
\caption{Evolution of the AGN fraction in clusters ($\lambda >20$) and groups ($\lambda <20$) for high redshift ($z>0.5$, blue diamonds), low redshift ($z<0.5$, red triangles), and all redshift AGN (black crosses). Points have been slightly offset in the x direction for visual clarity.}
\label{fig:withrich}
\end{figure}

We now investigate how the fraction of cluster galaxies hosting AGN depends on cluster redshift and richness.  Our main results are summarized in Table 2 and Figures~\ref{fig:withz} and \ref{fig:withrich}.  First, we consider the AGN fraction as a function of redshift for three redshift bins of $0.1<z<0.4$, $0.4<z<0.7$, $0.7<z<1.0$.  For these bins and including clusters of all richnesses the AGN fraction increases by a factor of 8.3 between the lowest and the highest redshift bins, and the high redshift AGN fraction ($0.7<z<1.0$ ) is greater than the low redshift AGN fraction ($0.1<z<0.4$) at $3.6\sigma$.  To separate the trend in redshift from any dependence of AGN fraction on cluster mass, we also look at the AGN fraction for low ($5<\lambda < 20$) and high richness ($\lambda > 20$) clusters separately.  A $\lambda=20$ cluster has a mass of $M_{500} \sim 10^{14} M_{\odot}$ while a $\lambda=5$ cluster has a mass of $M_{500} \sim 2 \times 10^{13} M_{\odot}$ \citep{2016arXiv160100621R}.  In terms of velocity dispersion, $\lambda=20$, the boundary between our low and high mass bins, corresponds to a cluster with $\sigma_v \sim 500$ km s$^{-1}$ \citep{2016arXiv160105773F}; this corresponds well to the division between group and cluster scale systems in \cite{2009ApJ...707.1691A}.  
The steep increase in AGN fraction with redshift is also present in the low and high-richness systems separately with a significance better than $2 \sigma$ for the groups and better than $3\sigma$ for the clusters.  Thus the trend of a larger fraction of cluster galaxies hosting X-ray AGN at higher redshifts holds regardless of cluster mass.

\begin{table*}
\centering
    \begin{tabular}{| p{2cm} | l | l | l | | p{2cm} | p{2cm} | p{2cm} |} \hline
    Richness Cut & Redshift Bin & AGN Fraction & $\sigma$ & $N_{\textrm{AGN}}$ & $N_{\textrm{gal}}$ & $N_{\textrm{clus}}$ \\ \hline
    $\lambda > 5$ & $0.1-0.4$ & 0.0011 & 0.0007 & 2 & 1611 & 116\\
    	 & $0.4-0.7$ & 0.0026 & 0.0011 & 8 & 2072 & 180  \\
     		  & $0.7-1.0$ & 0.0091 & 0.0021 & 23 & 1946 & 136  \\ \hline
     	   $\lambda > 20$ &$0.1-0.4$ & 0.0016 & 0.0010 & 2 & 834 & 22\\
     		   &$0.4-0.7$ & 0.0021 & 0.0011 & 3 & 767 & 29 \\
    			   &$0.7-1.0$ & 0.0109 & 0.0027 & 4 & 398 & 11 \\ \hline
                $5< \lambda < 20$ &$0.1-0.4$ & 0 & -- & 0 & 777 & 94 \\
     		   &$0.4-0.7$ & 0.0031&  0.0020 & 5 & 1305 & 151 \\
      			 &$0.7-1.0$ & 0.0081 & 0.0028 & 19 & 1548 & 125\\
        \hline
               
       	\end{tabular}
	\caption{AGN fraction results.  Columns 3 and 4 list the AGN fraction and $1 \sigma$ uncertainty in three redshift bins ($0.1<z<0.4$, $0.4<z<0.7$, and $0.7<z<1.0$) for varying cluster richness cuts. Columns 5--7 give the number of AGN, number of galaxies, and number of clusters in each bin. Note that these raw numbers have not been adjusted by the galaxy membership probabilities. The membership probabilities have been correctly accounted for in calculating the AGN fractions.}
	\label{agnfrac}
\end{table*}

\begin{figure}
\includegraphics[scale=0.5]{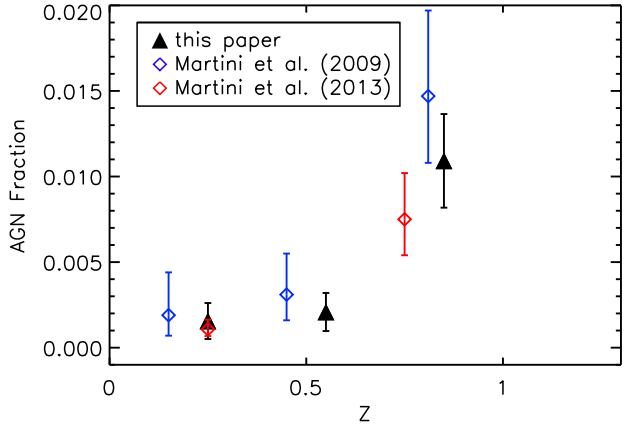}
\caption{Comparison of the evolution of the AGN fraction in clusters with redshift in our sample and in previous work. For all samples only AGN with X-ray luminosity of $L_{X}>10^{43}$ erg s$^{-1}$ are included. The results of this paper are shown in black, for three separate redshift bins of  $0.1<z<0.4, 0.4<z<0.7, 0.7<z<1.0$ (black filled triangles). Also shown are the AGN fraction for \citet{2009ApJ...701...66M} for z-bins of $0.05<z<0.3, 0.3<z<0.6, 0.6<z<1.3$ (blue diamonds) and the results of \citet{2013ApJ...768....1M} for $z<0.5$ and $0.5<z<1.0$ (red diamonds).  Here we show the results for our sample for clusters with a richness threshold of $\lambda > 20$ as the samples of \citet{2009ApJ...701...66M, 2013ApJ...768....1M} include primarily more massive systems.}
\label{fig:martini}
\end{figure}

As can be seen in Figure~\ref{fig:withz}, most of the evolution in the AGN fraction occurs for clusters at the highest redshifts in our sample with little or no evolution between the lowest redshift and intermediate redshift bins.  The increase in cluster AGN activity at increased redshift is similar to previous results \citep{2007ApJ...664L...9E, 2009ApJ...701...66M, 2013ApJ...768....1M}.  For example, \cite{2009ApJ...701...66M} find an increase by a factor of 7--8 in the cluster AGN fraction over a similar, though slightly broader redshift range ($0.05 < z < 1.3$) with most of the increase stemming from clusters at $z>0.6$.  This study employs similar X-ray luminosity ($L_X > 10^{43}$ erg s$^{-1}$, 2-10 keV) and optical galaxy magnitude cuts ($M_R^{\ast}(z) + 1$) as our work.  Figure~\ref{fig:martini} compares our results on the AGN fraction evolution to those of \cite{2009ApJ...701...66M} and \cite{2013ApJ...768....1M}; here we plot the AGN fractions only for higher-richness, $\lambda > 20$ redMaPPer clusters as the samples of \cite{2009ApJ...701...66M, 2013ApJ...768....1M} include primarily relatively massive systems.  Within the uncertainties, our AGN fractions are in excellent agreement with these previous studies despite the different selection of clusters and cluster galaxies.  Even with the very small area and thus limited cluster sample from DES SV the uncertainties on our AGN fractions are comparable to or better than previous results. Upcoming Year 1 through Year 3 DES cluster samples will allow us to trace the evolution in cluster AGN populations with unprecedented precision. An additional advantage of the current study is the uniform cluster and cluster member selection afforded by the DES data as well as the extension to lower richness/mass systems compared to previous work.

Table 2 and Figure~\ref{fig:withz} also show that the AGN fractions in each redshift bin for our sample are comparable for high and low richness clusters, which implies that the fraction of galaxies hosting AGN is not heavily dependent on the cluster mass. This is evident in Figure~\ref{fig:withrich}, which shows the AGN fraction for high ($\lambda >20$) and low richness ($5< \lambda <20$) clusters at all redshifts and separately at low ($z<0.5$) and high ($z>0.5$) redshifts.  There is no statistically significant difference in the AGN fractions in low and high richness systems at any redshift.  Table 3 lists the AGN fractions for high ($\lambda >20$) and low richness ($5< \lambda <20$) clusters in our lowest redshift bin ($0.1<z<0.4$) including AGN down to $L_X > 10^{42}$ erg s$^{-1}$.  While the AGN fraction in low richness systems seen here is mildly higher (by $\sim$ 40\%) this difference is not significant given the uncertainties.

\begin{table*}
\centering
    \begin{tabular}{| p{2cm} | l | l | l | | p{2cm} | p{2cm} | p{2cm} |} \hline
    Richness Cut & Redshift Bin & AGN Fraction & $\sigma$ & $N_{\textrm{AGN}}$ & $N_{\textrm{gal}}$ & $N_{\textrm{clus}}$ \\ \hline
     	   $\lambda > 20$ &$0.1-0.4$ & 0.0069 & 0.0021 & 8 & 688 & 22\\ \hline
                $5< \lambda < 20$ &$0.1-0.4$ & 0.0096 & 0.0042 & 9 & 640 & 94 \\
        \hline
               
       	\end{tabular}
	\caption{AGN fraction at low redshifts ($0.1<z<0.4$) including sources down to lower X-ray luminosities of $L_X > 10^{42}$ erg s$^{-1}$; all other cuts are the same as in Table 2.  Columns 3 and 4 list the AGN fraction and $1 \sigma$ uncertainty for two bins in cluster richness. Columns 5--7 give the number of AGN, number of galaxies, and number of clusters in each bin. Note that these raw numbers have not been adjusted by the galaxy membership probabilities. The membership probabilities have been correctly accounted for in calculating the AGN fractions.}
	\label{agnfrac}
\end{table*}

The lack of an observed trend in AGN activity with cluster mass is contrary to previous observations claiming a higher AGN fraction in lower mass systems \citep{2008ApJ...682..803S, 2009ApJ...707.1691A, 2014ApJ...790...43O}.  For example, \cite{2009ApJ...707.1691A} find that groups at very low redshifts ($0.02<z<0.06$) have an AGN fraction a factor of two higher than clusters at mild significance (85\%), while \cite{2014ApJ...790...43O} finds a similar factor of two difference in the AGN fractions of groups and clusters at higher redshifts ($z \sim 0.7$).  A mild, factor of two difference in AGN fraction between lower and higher mass systems is allowable within our errors bars, but there are also major differences between these previous studies and ours.  \cite{2009ApJ...707.1691A} considers groups at lower redshifts than our sample and AGN with X-ray luminosities up to two orders of magnitude fainter than we use here ($L_X > 10^{41}$ erg cm$^{-2}$ s$^{-1}$); \cite{2014ApJ...790...43O} also considers fainter AGN at high redshifts than our study ($L_X > 10^{42}$ erg cm$^{-2}$ s$^{-1}$), and both papers sample X-ray selected groups.   X-ray selection requires a dense, collapsed system such that there is detectable emission from a hot intragroup medium and may not select systems with similar dynamical states to selection based on galaxy content, particularly at the low-mass end.  For example, in a small sample of low-redshift, low-mass groups selected via optical spectroscopy, \cite{2007ApJ...654L.115S} find indications of a high fraction of optical emission-line selected AGN and a lower fraction of X-ray AGN compared to X-ray luminous groups.

If galaxy-galaxy mergers contribute significantly to the presence of AGN activity, one might expect a larger AGN fraction in group scale systems where the smaller velocity dispersions of galaxy members make them more susceptible to mergers.  In particular, a higher merger rate could significantly enhance the number of high-luminosity AGN \citep[e.g.][]{2008ApJS..175..356H}.  In contrast, our present results, if verified by a larger sample, would suggest that galaxy mergers are not significant enough to enhance AGN activity in these systems \citep[see also][]{2014MNRAS.439.3342V}.  With the order of magnitude larger cluster samples soon to be available from DES Years 1--3, we expect to be able to conclusively test the correlation of AGN activity with cluster mass.  Given the overlap with existing Chandra data we expect to sample a factor of 10--20 more clusters with these data.

As discussed in \S2, cluster members in our study are selected to lie on or near the red-sequence, and therefore our AGN sample does not include AGN hosted by bluer cluster galaxies.  Interestingly, Figure 3 shows that the AGN fractions we measure are consistent within the uncertainties with previous studies where cluster membership was primarily determined based on spectroscopy.  X-ray detected AGN in the field tend to be detected in redder, more massive galaxies than the total galaxy population at the same redshifts; however, compared to galaxies of similar stellar mass, X-ray AGN hosts have a similar distribution of colors \citep[e.g.][]{2015A&ARv..23....1B, 2012MNRAS.427.3103B, 2013ApJ...763...59R}.  The fraction of blue versus red galaxies hosting X-ray AGN in clusters is less clear as there is little data in the literature at similar redshifts to our study.  At somewhat higher redshifts than those considered here ($z=1-1.5$), \cite{2013ApJ...768....1M} find that about half of their 11 X-ray detected, cluster AGN have blue host galaxy colors.  In general, high-redshift clusters contain a higher fraction of blue galaxies than low-redshift clusters with the blue fraction reaching roughly 50\% at $z=1$ \citep[e.g.][]{2016arXiv160400988H}, and in fact, the AGN fraction and the fraction of star-forming galaxies in clusters increase with redshift at similar rates \citep{2013ApJ...768....1M}.  If the fraction of blue cluster galaxies is similar to the fraction of cluster AGN in blue galaxies then the total AGN fraction will not change substantially by including these sources, which is at least qualitatively consistent with what we find.  What is clear is that the fraction of red cluster galaxies hosting AGN increases substantially with redshift, and both the AGN fractions and the rate of evolution we find are similar to previous studies.

\section{Conclusions}

We study the X-ray luminous AGN populations in clusters of galaxies observed in DES Science Verification.  Specifically, we search for X-ray point sources in archival Chandra data associated with cluster member galaxies from the DES SVA1 expanded cluster catalog for clusters over a large richness/mass range.  Using a sample of 33 X-ray detected AGN associated with red-sequence cluster member galaxies meeting our X-ray and optical luminosity completeness cuts ($L_X > 10^{43}$ erg s$^{-1}$ and host galaxy luminosity brighter than $0.5 L_{\ast}$ in z-band), we find that the fraction of cluster galaxies hosting AGN increases strongly with redshift.  The AGN fraction increases sharply to about 1\% for $z > 0.7$ with mild to no evolution at lower redshifts.  Our results are in good agreement with previous studies using smaller cluster samples \citep{2007ApJ...664L...9E, 2009ApJ...701...66M, 2013ApJ...768....1M}.  The increase in AGN activity in clusters mirrors the increase in star formation in cluster galaxies over the same redshift range \citep[e.g.][]{1984ApJ...285..426B, 2006ApJ...642..188P, 2008ApJ...685L.113S, 2009ApJ...704..126H, 2014MNRAS.445.2725E}.  We do not, however, find any significant trend in AGN fraction with cluster richness in any redshift bin.  

The strong increase in the cluster AGN population with redshift is a systematic effect that may have important implications for cluster selection in X-ray, e.g. for the future eROSITA mission \citep{2012arXiv1209.3114M}, and via the Sunyaev-ZelÕdovich Effect for radio-loud AGN \citep{2016arXiv160505329G, 2015MNRAS.452.2353K}.  The data employed here represent only $\sim4$\% of the area that will be covered by the full DES survey; with the full survey we will thus be able to trace with high accuracy cluster AGN populations and their correlation to mass and redshift.  

\section*{Acknowledgments}

EB would like to acknowledge support from the Julia Packard SRE and the UCSC STEM Diversity Program.  This material is based upon work supported by the U.S. Department of Energy, Office of Science, Office of High Energy Physics program under Award Number DE-SC-0013541.  This research has made use of data obtained from the {\em Chandra} Data Archive and software provided by the {\em Chandra} X-ray Center (CXC) in the application package CIAO.

Funding for the DES Projects has been provided by the U.S. Department of Energy, the U.S. National Science Foundation, the Ministry of Science and Education of Spain, 
the Science and Technology Facilities Council of the United Kingdom, the Higher Education Funding Council for England, the National Center for Supercomputing 
Applications at the University of Illinois at Urbana-Champaign, the Kavli Institute of Cosmological Physics at the University of Chicago, 
the Center for Cosmology and Astro-Particle Physics at the Ohio State University,
the Mitchell Institute for Fundamental Physics and Astronomy at Texas A\&M University, Financiadora de Estudos e Projetos, 
Funda{\c c}{\~a}o Carlos Chagas Filho de Amparo {\`a} Pesquisa do Estado do Rio de Janeiro, Conselho Nacional de Desenvolvimento Cient{\'i}fico e Tecnol{\'o}gico and 
the Minist{\'e}rio da Ci{\^e}ncia, Tecnologia e Inova{\c c}{\~a}o, the Deutsche Forschungsgemeinschaft and the Collaborating Institutions in the Dark Energy Survey. 

The Collaborating Institutions are Argonne National Laboratory, the University of California at Santa Cruz, the University of Cambridge, Centro de Investigaciones Energ{\'e}ticas, 
Medioambientales y Tecnol{\'o}gicas-Madrid, the University of Chicago, University College London, the DES-Brazil Consortium, the University of Edinburgh, 
the Eidgen{\"o}ssische Technische Hochschule (ETH) Z{\"u}rich, 
Fermi National Accelerator Laboratory, the University of Illinois at Urbana-Champaign, the Institut de Ci{\`e}ncies de l'Espai (IEEC/CSIC), 
the Institut de F{\'i}sica d'Altes Energies, Lawrence Berkeley National Laboratory, the Ludwig-Maximilians Universit{\"a}t M{\"u}nchen and the associated Excellence Cluster Universe, 
the University of Michigan, the National Optical Astronomy Observatory, the University of Nottingham, The Ohio State University, the University of Pennsylvania, the University of Portsmouth, 
SLAC National Accelerator Laboratory, Stanford University, the University of Sussex, Texas A\&M University, and the OzDES Membership Consortium.

The DES data management system is supported by the National Science Foundation under Grant Number AST-1138766.
The DES participants from Spanish institutions are partially supported by MINECO under grants AYA2012-39559, ESP2013-48274, FPA2013-47986, and Centro de Excelencia Severo Ochoa SEV-2012-0234.
Research leading to these results has received funding from the European Research Council under the European UnionÕs Seventh Framework Programme (FP7/2007-2013) including ERC grant agreements 
 240672, 291329, and 306478.
 
 We are grateful for the extraordinary contributions of our CTIO colleagues and the DECam Construction, Commissioning and Science Verification
teams in achieving the excellent instrument and telescope conditions that have made this work possible.  The success of this project also 
relies critically on the expertise and dedication of the DES Data Management group.

\bibliography{agn}

%\bibliography{}
\end{document}